\documentclass[prb,twocolumn,showpacs,graphicx]{revtex4}
\usepackage{amsfonts}
\usepackage{amssymb}
\usepackage{amsmath}
\usepackage{graphicx}

\begin{document}

\title{Gap opening in graphene by simple periodic inhomogeneous strain}
\author{I.I. Naumov and A.M. Bratkovsky}
\affiliation{Hewlett-Packard Laboratories, 1501 Page Mill Road, Palo Alto, California
94304}
\date{\today }
\keywords{}

\begin{abstract}
Using ab-initio methods, we show first that the \emph{uniform} deformation
either leaves graphene (semi)metallic or opens up a small gap yet only
beyond its mechanical breaking point, contrary to claims in the literature
based on tight-binding (TB)\ calculations. It is possible, however, to open
up a global gap by a \emph{sine-like} inhomogeneous deformation applied
along any direction but the armchair one, with the largest gap along the
zigzag direction ($\sim 1.0$ eV) and without any electrostatic gating. The
gap opening has a threshold character with very sharp rise when the ratio of
the amplitude $A$ and the period of the sine wave deformation $\lambda $
exceeds ($A/\lambda )_{c}\sim 0.1$ and the inversion symmetry is preserved,
while it is threshold-less when the symmetry is broken. The gap opening
occurs in graphene mesh on boron-nitride substrate.
\end{abstract}

\pacs{73.22.Pr,  81.05.ue, 62.25.-g}
\maketitle

\section{Introduction}

Graphene, a hexagonally packed single layer of carbon atoms \cite{novomos},
is considered a strong candidate for post-silicon electronic devices. To
meet this expectation, however, it is crucial to be able to open up a band
gap at the Fermi level, which exists in silicon and is necessary for a
switchable device. Since graphene has outstanding mechanical properties and
capable of sustaining huge atomic distortions \cite{lee,kim} up to about
27\% \cite{liu07} it becomes very interesting to use strain as a tool to
introduce a band gap in graphene.

The idea of strain-induced band gap has been theoretically explored by
Pereira \emph{et al.} \cite{pereira09} within the standard tight-binding
(TB) approximation. The gap opening process involves merging of two
inequivalent Dirac points, $\boldsymbol{D}$ and $-\boldsymbol{D.}$\cite%
{pereira09,mont08,montambaux} Such a process requires some $\sim 23\%$ of
stretching along the zigzag chains \cite{pereira09}. The subsequent to Ref.%
\onlinecite{pereira09} and more realistic \emph{ab-initio} calculations \cite%
{ni1, ni2,choi} confirmed that the strain along a \emph{zigzag} direction is
indeed the most effective in annihilating the Dirac points. However, the
ab-initio methods predicted even higher critical tensile strain ($\approx
27\%$) than that obtained within the tight-binding model. Moreover, it was
discovered that annihilation of the Dirac points does not \emph{automatically%
} lead to a gap opening, because a $\sigma ^{\ast }$ conduction band quickly
moves down towards the Fermi level with the strain. As a result, when the
Dirac points merge, the energy gap either does not open at all \cite{ni2},
or opens by a tiny 45 meV \cite{choi}. Thus, the results \cite%
{pereira09,ni2,choi} suggest that a tangible gap in graphene is unlikely
under uniaxial tensions up to the graphene failure strain of 25-27\% \cite%
{lee,liu07,cadelano}. The papers \cite{pereira09,ni1,ni2,choi} considered a
tension accompanied by a Poisson's contraction. Using a TB approach, Cocco
\emph{et al.} \cite{cadelano2} have tried a uniform \emph{pure shear} strain
$\epsilon _{xy}=\epsilon _{yx}\neq 0$ associated with the elastic constant $%
C_{44}$. They claimed a moderate critical value of $16\%$ and expected it to
lower down to $12\%$ if used along with the uniaxial strain in the armchair
direction.

The idea of using inhomogeneous strain fields $u$ with $\nabla _{i}u_{k}\neq
0$ has also been studied extensively bearing on TB-derived notion of a
pseudo-magnetic field $B\propto \nabla u$. \cite{pacoNP10,levy10,low11} It
was claimed that two-dimensional corrugations with triangular symmetry open
up a gap without a threshold \cite{pacoNP10}. The local gaps in graphene
`bubbles' have been reported in Ref.~\onlinecite{levy10}, while it was
speculated that the one-dimensional corrugations (`wrinkles') that are
frequently observed in a suspended graphene could open up the gap but only
in combination with alternating electrostatic gating correlated with the
`wrinkles' \cite{low11}.

Here, we pursue two goals with the use of ab-initio calculations applied
similarly to our earlier study of a flexo-electric [polarization induced by $%
\partial _{x}u_{z}\partial _{y}u_{z}$, $\left( \partial _{x}u_{z}\partial
_{y}u_{z}\right) ^{2}$ and $\left( \partial _{x}\partial _{y}u_{z}\right)
^{2}$] `sister' system B-N monolayer \cite{bn}. First, we show that contrary
to Ref.~\onlinecite{cadelano2} graphene's band structure remains gapless
under \emph{any uniform strains} not exceeding its breaking point. Secondly,
we demonstrate that the energy gap can be nevertheless induced mechanically
by applying realistic \emph{nonuniform} sine-wave deformations (similar to
the `wrinkles') in all but the armchair directions, and \emph{without} any
periodic gating \cite{low11}. The qualitative difference with TB-derived
continuous pseudo-magnetic models (see Ref.~\onlinecite{low11} and
references therein) appears to be an account for strong $pp\sigma $ overlap
between neighboring $p_{z}$ orbitals on Carbon atoms appearing at any
flexing with $\nabla u\neq 0,$ in addition to the standard $pp\pi $ overlap,
the only one retained in TB\ models of graphene. Since the rehybridization
responsible for the gap opening is local, the results do not depend on
standard procedure of employing supercell for the band structure
calculations and atomic relaxation. Another feature missing in the
continuous flexo-`magnetic' models is that they do not distinguish between
flexing that is preserving an inversion symmetry versus a symmetry breaking
one.

\section{Homogeneous and 1D periodic deformations of Graphene}

We begin with defining the homogeneous and inhomogeneous (sinusoidal)
deformations of graphene. Select the unit vectors for real lattices of an
undistorted 2D graphene as $\boldsymbol{a}_{1}=\frac{a}{2}\left( \sqrt{3}%
,-1,\right) ,$ $\boldsymbol{a}_{2}=\frac{a}{2}\left( \sqrt{3},1\right) $
with the reciprocal lattice vectors $\boldsymbol{G}_{1}=\frac{2\pi }{\sqrt{3}%
a}\left( 1,-\sqrt{3}\right) ,$ $\boldsymbol{G}_{2}=\frac{2\pi }{\sqrt{3}a}%
\left( 1,\sqrt{3}\right) $. They are chosen in such a way that the $x-$axis
is along the armchair direction $(\boldsymbol{a}_{1}+\boldsymbol{a}_{2})$,
while the $y$ axis is along the zigzag direction $(-\boldsymbol{a}_{1}+%
\boldsymbol{a}_{2})$. Let us consider a general homogeneous deformation of a
graphene sheet. Excluding overall translation and rotations, such a change
can be described by the displacements:
\begin{equation}
{\binom{\delta x}{\delta y}}=\left(
\begin{array}{cc}
\epsilon _{0}+\eta & \gamma \\
\gamma & \epsilon _{0}-\eta%
\end{array}%
\right) {\binom{x}{y},}  \label{eq:dx}
\end{equation}
where $\epsilon _{0}$ corresponds to an isotropic distortion (dilatation),
while the parameters $\eta $ and $\gamma $ describe the pure independent
shear strains. By rotating the frame around the $z$ axis through an angle $%
\phi $ given by $\cos \left( 2\phi \right) =\eta /\sqrt{\eta ^{2}+\gamma ^{2}%
}$, one can diagonalize the stress tensor in (\ref{eq:dx}):

\begin{equation}
{\binom{\delta x^{\prime }}{\delta y^{\prime }}}=\left(
\begin{array}{cc}
\epsilon _{0}+\tau & 0 \\
0 & \epsilon _{0}-\tau%
\end{array}%
\right) {\binom{x^{\prime }}{y^{\prime }}}\equiv \left(
\begin{array}{cc}
\epsilon _{11} & 0 \\
0 & \epsilon _{22}%
\end{array}%
\right) {\binom{x^{\prime }}{y^{\prime }}},  \label{eq:dx1}
\end{equation}%
where $\tau =\sqrt{\eta ^{2}+\gamma ^{2}},$ $\epsilon _{11}=\epsilon
_{0}+\tau $ and $\epsilon _{22}=\epsilon _{0}-\tau $.

For \emph{inhomogeneous} deformations, it is sufficient to consider a
periodic out-of-plane atomic displacements $u_{z}(\boldsymbol{r})=A\sin (%
\boldsymbol{k\cdot r}+\varphi )$, where $\boldsymbol{k}$ is the undulation
wave vector, $\boldsymbol{r}$ is the in-plane vector with the origin at the
lattice \emph{inversion center}, and $\varphi $ some phase. To set up the
supercell calculations, we use a periodically and commensurately distorted
graphene sheets, the corrugation wave vector is $\boldsymbol{k}=2\pi
\boldsymbol{e}/{\lambda }$, where the unit vector $\boldsymbol{e}$ and the
wavelength $\lambda $ are given by $\boldsymbol{e}=\boldsymbol{\lambda }%
/\lambda $, $\boldsymbol{\lambda }(n,m)=n\boldsymbol{a}_{1}+m\boldsymbol{a}%
_{2}$, $\lambda =a\sqrt{n^{2}+nm+m^{2}}$, where $n$ and $m$ are integers, $|%
\boldsymbol{a}_{1}|=|\boldsymbol{a}_{2}|=a$, cf. \cite{saito,note1}. The
corrugation with the period $\lambda (n,m)$ leads to a rectangular supercell
with translational vectors $\boldsymbol{\lambda }(n,m)$ and $\boldsymbol{T}=N%
\boldsymbol{a}_{1}+M\boldsymbol{a}_{2}$, $\boldsymbol{T}\perp \boldsymbol{%
\lambda }$, $N=(2m+n)/d_{R},\,M=-(2n+m)/d_{R}$, where $d_{R}$ is the
greatest common divisor of $2m+n$ and $2n+m$.\cite{marconcini} The number of
graphene unit cells per supercell $\boldsymbol{\lambda }\times \boldsymbol{T}
$ is $N{_{g}}=2\lambda ^{2}/(a^{2}d_{R})$.

For the undeformed graphene, the two inequivalent Dirac points are at $(2%
\boldsymbol{G}_{1}+\boldsymbol{G}_{2})/3$ and $(\boldsymbol{G}_{1}+2%
\boldsymbol{G}_{2})/3$, respectively, at the two corners $\mathbf{K}$ and $%
\mathbf{K}^{\prime }$ of the first Brillouin zone (BZ) \cite{montambaux}. It
is easy to find the positions of these points inside the rectangular BZ of
the supercell (when $A\rightarrow $ 0) using a so-called zone-folding
technique. Namely, $\boldsymbol{k}=2\pi \boldsymbol{e}/{\lambda }$ is of one
of the reciprocal vectors of the supercell, equal to $(-M\boldsymbol{G}_{1}+N%
\boldsymbol{G}_{2})/N{_{g}}$. The other reciprocal vector is $\boldsymbol{s}%
=(m\boldsymbol{G}_{1}-n\boldsymbol{G}_{2})/N{_{g}}$, and we find $\mathbf{K}%
=(2n+m)\boldsymbol{k}/3+(2N+M)\boldsymbol{s}/3,\,\mathbf{K}^{\prime }=(n+2m)%
\boldsymbol{k}/3+(N+2M)\boldsymbol{s}/3.$\cite{marconcini} In particular,
for the corrugation along the zigzag direction with $\boldsymbol{\ \lambda }%
=(6,0)$, both the $\mathbf{K}$ and $\mathbf{K}^{\prime }$ points are
translated into the origin ($\Gamma -$point).

The numerical calculations for both the flat and the corrugated graphene
sheets have been performed using density functional theory implemented in
the ABINIT package \cite{abinit} A 16$\times $16$\times $1 Monkhorst-Pack $%
\boldsymbol{k}$-point grid \cite{monkhorst} has been used in the case of
flat graphene. Approximately the same k-point density was kept in going from
the planar to a corrugated graphene. The sheets have been simulated by a
slab-supercell approach with the inter-planar distances of $30a_{B}$ to
ensure negligible wave function overlap between the replica sheets. For the
plane-wave expansion of the valence and conduction band wave-functions, a
cutoff energy was chosen to be 80 Ry. The plane waves Troullier-Martins \cite%
{troullier} and Fritz-Haber-Institute \cite{fuchs} pseudopotentials have
been used for the calculations in the local density (LDA) and the
generalized gradient (GGA) approximations, respectively. In both cases,
Carbon 2s and 2p electrons have been considered as valence states.

\section{No gap in graphene subject to a homogenous deformation}

Since any uniform strain respects the inversion symmetry, it can remove the
Dirac point degeneracy only if it can force the Dirac points to merge, in
compliance with the Wigner-von Neumann theorem. Merging may only take place
at a point located at half the reciprocal lattice vector, $\boldsymbol{g}%
/2=(p\boldsymbol{G}_{1}+q\boldsymbol{G}_{2})/2,$ where $p,q$ are the
integers \cite{montambaux}. Obviously, the Dirac point can be moved with
regards to the reciprocal vectors only when the strain comprises the shear $%
\tau $. Therefore, below we will distinguish three important cases of the
\emph{homogeneous} strain: (i) pure shear strain when $\epsilon _{0}=0,$ $%
\epsilon _{11}=\tau ,$ $\epsilon _{22}=-\tau $ in (\ref{eq:dx1}), (ii)
uniaxial tension accompanied by Poisson's transverse contraction, $\epsilon
_{0}=\tau (1+\sigma )(1-\sigma ),$ $\epsilon _{22}=-\epsilon _{11}\sigma $,
with $\sigma $ the Poisson coefficient, and (iii) pure uniaxial tension ($%
\epsilon _{0}=\tau ,$ $\epsilon _{11}=2\tau ,$ $\epsilon _{22}=0$).

\begin{figure}[h]
\begin{centering}
\includegraphics
[width=8.5cm]{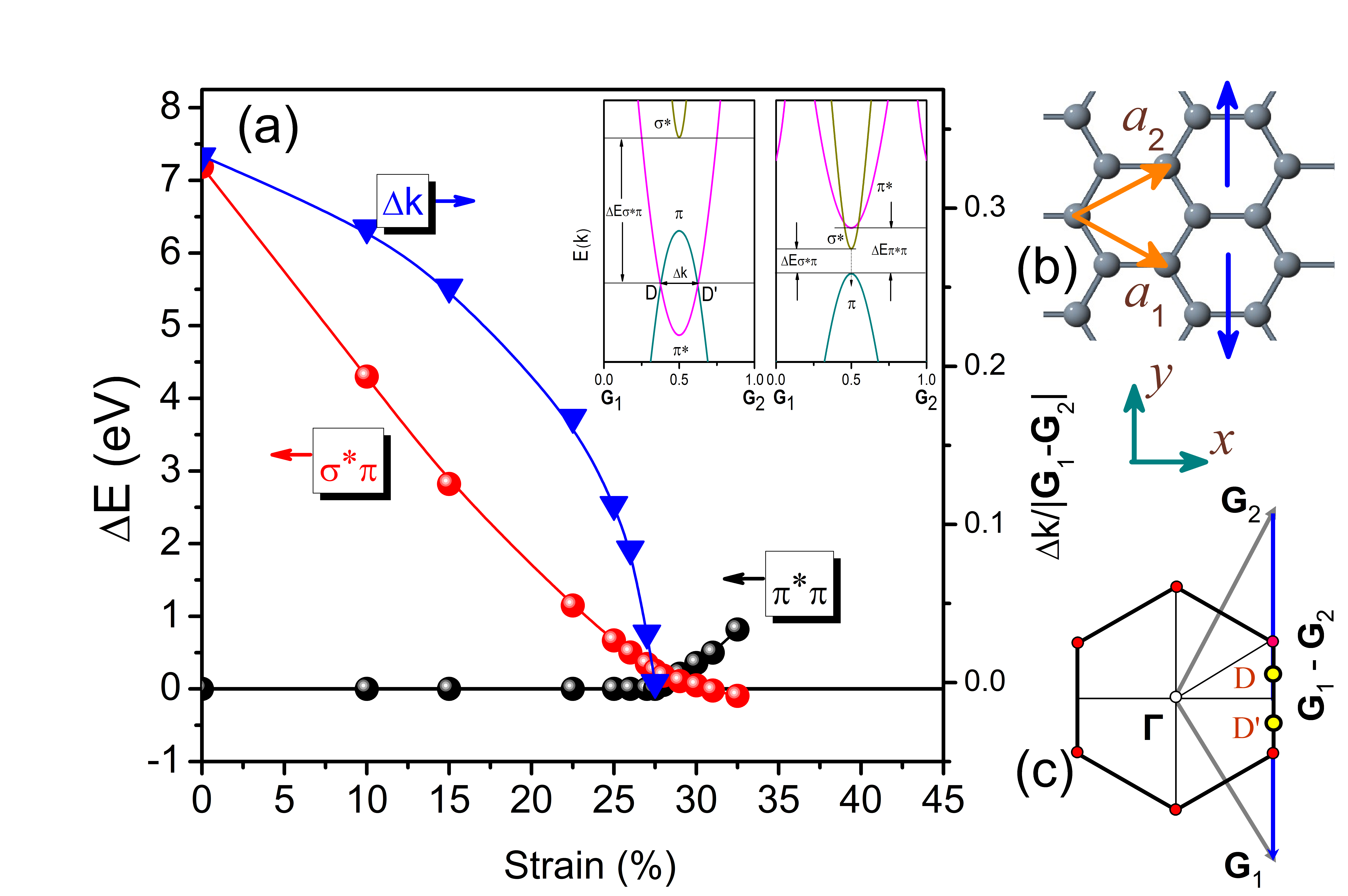}
\caption{(color online).
((a) The evolution of the gaps between
the valence $\pi $ and the conduction $\pi ^{\ast }$ and $\sigma ^{\ast }$
bands under uniaxial strain schematically shown in panel (b). The Dirac
points move along the straight trajectory in BZ before merging (c). The
bands along the merging direction from $\boldsymbol{G_{1}}$ to $\boldsymbol{%
G_{2}}$ are shown in the inset in (a) with the left (right) inset showing
the bands before (after) the merging of Dirac points.
}\label{fig:1S}
\end{centering}
\end{figure}

We have performed detailed analysis with account for full atomic relaxation
of the above cases (i)-(iii) exemplified by a graphene sheet under a tensile
strain $\epsilon _{yy}$ with varying conditions for a perpendicular strain.
In this geometry, the Dirac points merge in the case (i), when $\epsilon
_{yy}=-\epsilon _{xx}$, and the strain $\epsilon _{yy}$ reaches $23.5\%$,
while in the cases (ii) and (iii) for $\epsilon _{yy}$ exceeding $26.5$ and $%
27.5\%$, respectively. Since the rate of $\sigma ^{\ast }-$band lowering
increases in passing from (iii) to (ii) and then from to (ii) to (i), the
gap is unable to open up in the cases (i),(ii) at all. Although the gap does
open up in the case (iii), its maximal value is modest and does not exceed
100 meV since the antibonding $\sigma ^{\ast }-$band quickly moves down in
energy and closes the gap (see Fig.~\ref{fig:1S}, cf. \cite{ni2}).
\begin{figure}[h]
\begin{centering}
\includegraphics
[width=8.5cm]{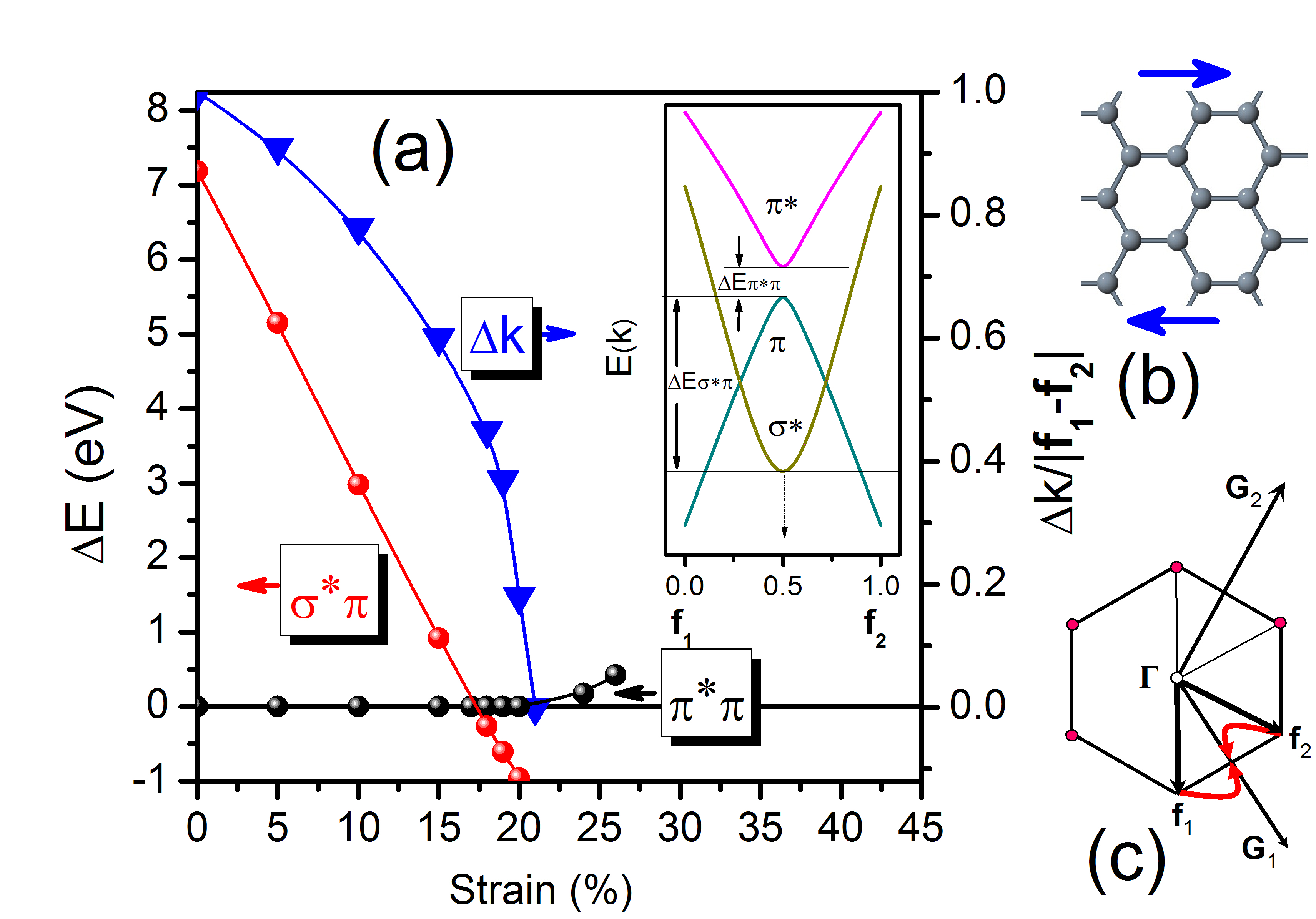}
\caption{(color online).
(a) The evolution of the gaps between
the valence $\pi $ and the conduction $\pi ^{\ast }$ and $\sigma ^{\ast }$
bands under pure shear strain shown schematically in panel (b). The Dirac
points move along complex trajectory in BZ before merging (c), where $%
\boldsymbol{f}_{1}=(\boldsymbol{G}_{1}-\boldsymbol{G}_{2})/3$, $\boldsymbol{f%
}_{2}=(2\boldsymbol{G}_{1}+\boldsymbol{G}_{2})/3$. The gap between $\pi
^{\ast }$ and $\pi $ states opens up, but not before the $\sigma ^{\ast
}-\pi $ gap closes (inset in (a)).
}\label{fig:2S}
\end{centering}
\end{figure}
The suppressing role of $\sigma ^{\ast }-$band can be traced back to the
fact that the uniaxial strain changes the bond lengths and bond angles
between C-C bonds along the stretching direction and perpendicular to it.
The $sp^{\mathrm{2}}$\ hybrid tends to split into two $sp_{\parallel }$\
hybrids and the perpendicular to the stretching direction $p_{\perp }$\
chain states get progressively weaker coupled between themselves. As a
result, the $\sigma ^{\ast }-$band derived from $p_{\perp }$ states quickly
moves down at the point $\boldsymbol{k}/2$ \cite{harrbk} thus fully
suppressing the gap.

The above results are obtained in the local density approximation (LDA)\ for
the exchange-correlation energy. Using the generalized gradient
approximation (GGA) instead does not change the results significantly, even
though GGA usually favors gap opening. For instance, in the case (ii) GGA
gives the critical value for the gap opening $\epsilon _{yy}=25.3\%$ with
the maximal gap value of about 100 meV. Keeping this in mind, below we
present only the LDA results.

Scanning all possible deformations (i)-(iii) corresponding to different
angles $\phi $ shows that the gap opening is most problematic in the case of
deformations involving stretching along the armchair directions. Here, the $%
\boldsymbol{D}$, $\boldsymbol{D}^{\prime }$ points should move the \emph{%
longest} distance in $\boldsymbol{k}$-space to meet each other, and the
latter can happen only well above the mechanical breaking strain. At the
same time, the \emph{\ pure shear} deformation conjugated to $C_{44}$
elastic modulus ($\phi =\pi /4,$ $\epsilon _{11}=-\epsilon _{22})$ was found
to be the most favorable for the Dirac points merging, in agreement with
Ref.~\onlinecite{cadelano2}. Evolution of the band structure under such a
strain is shown in Fig.~\ref{fig:2S}. In this case, the conduction $\sigma
^{\ast }- $band minimum moves down (inset in Fig.~\ref{fig:2S}a) and
intercepts merging of two Dirac points when shear strain reaches 17\%, while
the merging occurs only when graphene sheet is strained by about 22\%. One
concludes that the homogeneous strain is unable to open up a gap in
graphene, contrary to the claim in Ref.~\onlinecite{cadelano2}.

\section{Gap in graphene subject to two types of 1D corrugation}

\begin{figure}[h]
\begin{centering}
\includegraphics
[width=8.5cm]{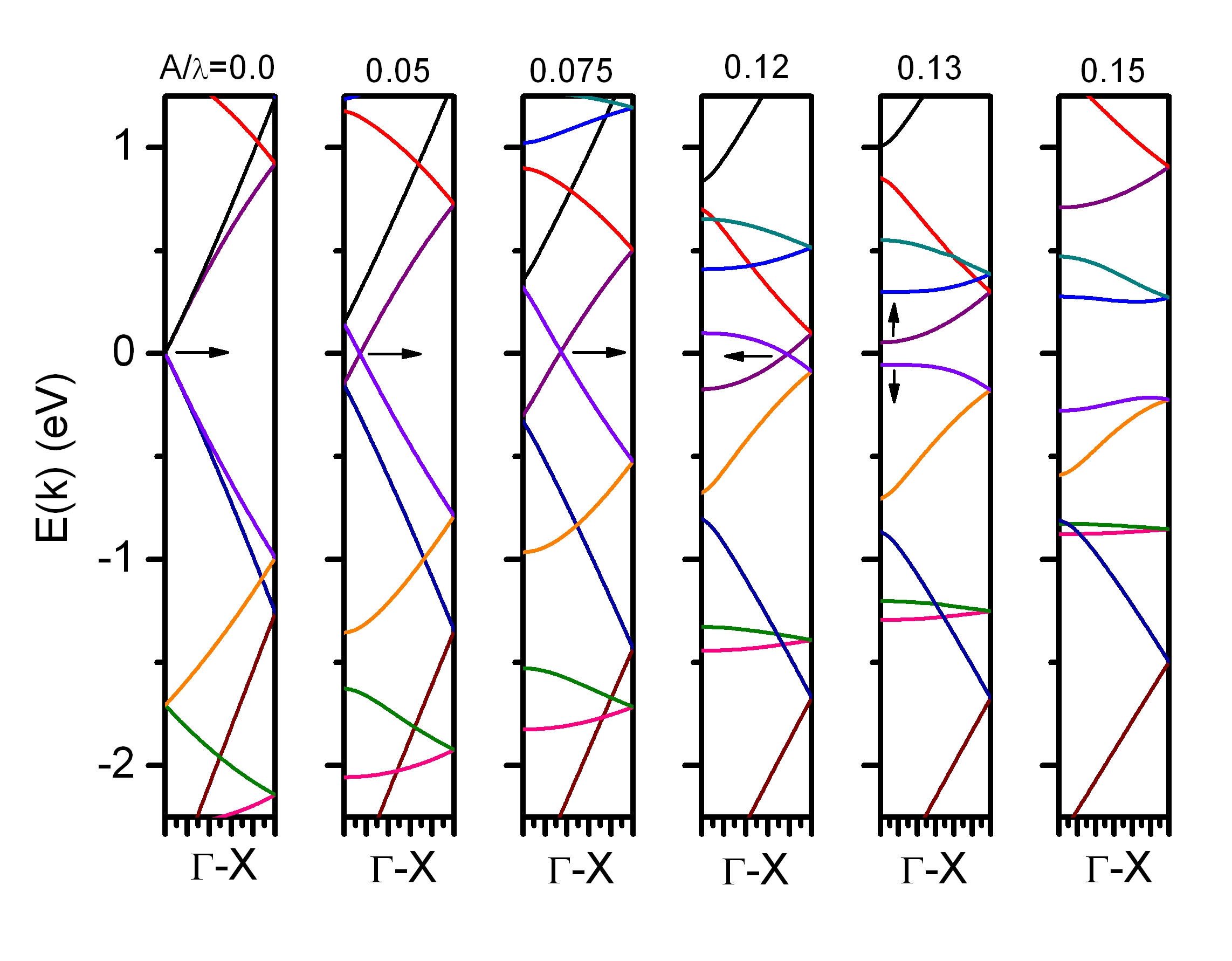}
\caption{(color online).
Bands in graphene ($\Gamma-X$ direction) corrugated along
the zigzag direction with $\boldsymbol{ \lambda} =(6,0)$, amplitude to period ratio $A/\lambda= 0-0.15$
and the phase $\varphi = 0$ preserving the inversion symmetry.
The Dirac ponts merge and the gap opens up when the corrugation is slightly smaller than
$A/\lambda=0.13$.
}\label{fig:corrzig}
\end{centering}
\end{figure}

Fortunately, the gap can be opened up in all but the armchair directions by
the corrugation, with the maximal gap of about 0.5 eV for the zigzag
direction. One can distinguish two qualitatively different regimes depending
on a phase of the corrugation $\varphi $: (a)\ $\varphi =0,$ the inversion
(and time reversal) symmetry is preserved, and (b)\ $\varphi \neq 0,$ the
inversion symmetry is generally broken. For the case (a), we have calculated
corrugation along the \emph{zigzag} direction $\boldsymbol{\lambda }=(6,0)$,
$\varphi =0$ in Fig.~\ref{fig:corrzig}. Such a perturbation is momentum
dependent preserving time reversal and inversion symmetry and therefore can
not automatically open up the gap in the graphene sheet \cite{manes07}
unless the Dirac points are merged at some point $\boldsymbol{g}/2.$ \cite%
{montambaux} In the case under consideration, the Dirac point moves on the
symmetry $\Gamma -X$ line in the folded BZ, where $X=\boldsymbol{k}/2,$ and
`bounces' off the $X$-point when $A/\lambda $ is close to $0.12$. Finally,
the Dirac points merge at the $\Gamma -$point. The gap opens up when $%
A/\lambda $ exceeds a critical value of $0.13$ and quickly reaches a
substantial size of 0.5 eV, as shown in Fig.~\ref{fig:corrzig} for $%
A/\lambda =0.15$ (see also Fig.~\ref{fig:GonBN}a). The wider maximal gap
relative to the case of a stretched flat lattice is simply due to the fact
that now the $\sigma ^{\ast }$ band is affected significantly less.

\begin{figure}[h]
\begin{centering}
\includegraphics
[width=8cm]{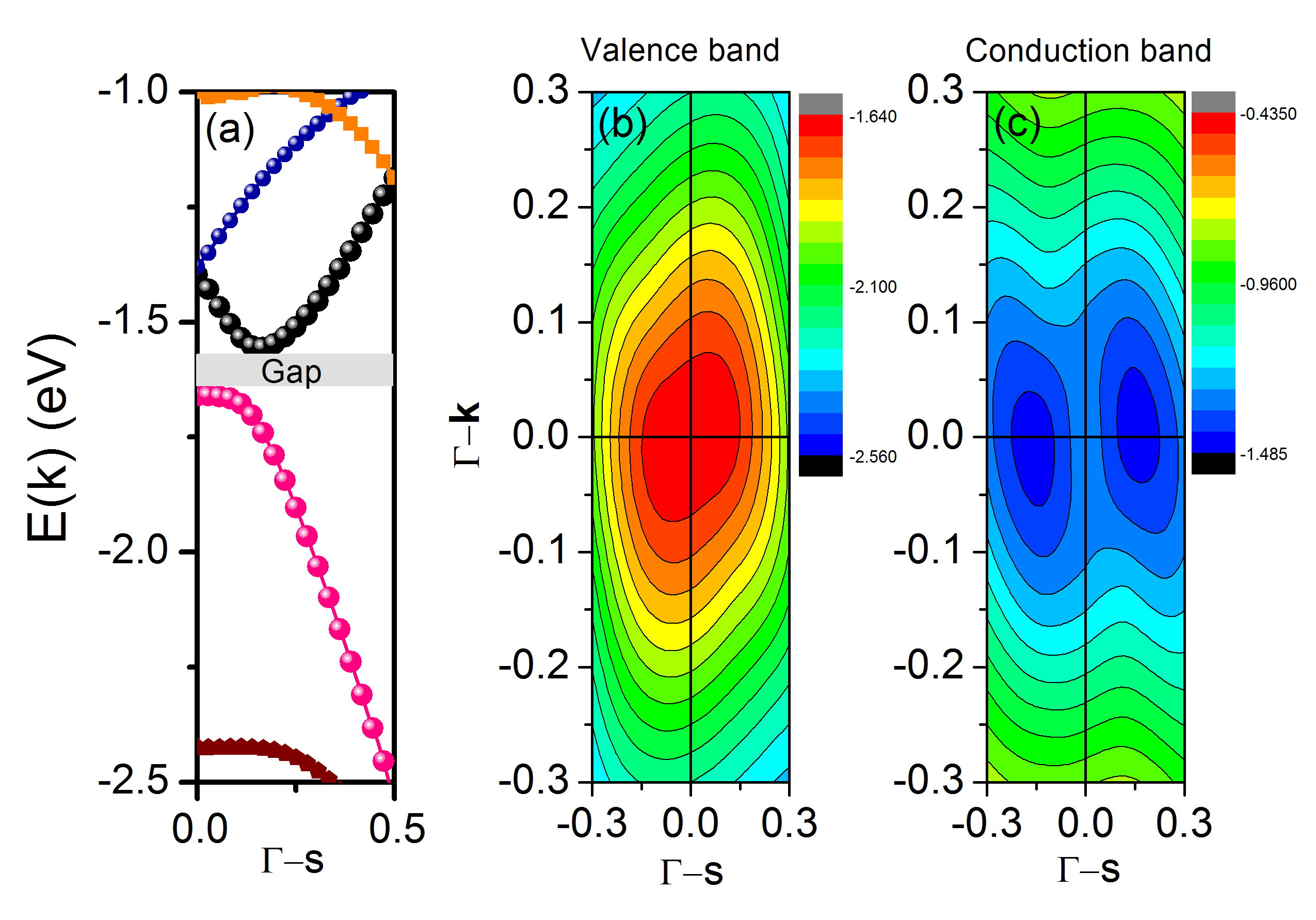}
\caption{(color online).
(a) Gap in graphene corrugated along the chiral direction $\boldsymbol {\lambda} =(-2,3),$
with $A/\lambda =0.14$ and  the phase  $\varphi  =3\pi/19$ breaking the inversion symmetry.
The Dirac cones detach at $A \neq 0$ and after that their remainders evolve
differently with corrugation  forming the indirect gap (b),(c).
}\label{fig:chir}
\end{centering}
\end{figure}

For possible applications, it is important to establish if the gap can be
opened up when corrugation runs in a general \emph{chiral direction} with an
arbitrary phase $\varphi$. Band structure of graphene subject to a
corrugation along a `chiral' direction, $\boldsymbol{\lambda }=(-2,3),$ $%
\varphi =3\pi /19,$ $A/\lambda =0.14$ is shown in Fig.~\ref{fig:chir}. Since
$\varphi \neq 0$ and the inversion symmetry is broken, the gap opens up at
any $A\neq 0.$ However, the gap is very small if the `pseudo' Dirac points
remain well inside the BZ, which is the case for small corrugations. When
the Dirac points approach the BZ boundary with increasing corrugation, the
minimum of the conduction and the maximum of the valence band move apart in $%
k$-space. Consequently, the initially small gap grows to (indirect) gap of
0.5 eV, Fig.~\ref{fig:chir}.

We should stress that in several aspects our results are in contrast with
the gap behavior claimed on the basis of the continuous flexo-`magnetic'
models \cite{low11, mikmidgap08,guinea}. First, the latter are oblivious of
the presence of the inversion symmetry or its lack thereof \cite{mikmidgap08}%
. As a result, they are missing the important increase in the gap size from
trivially expected minute value in the symmetry-breaking case to large
values when the remainders of the Dirac cones approach the BZ boundaries,
Fig.~\ref{fig:chir}. Second, even if the inversion symmetry is preserved,
the gap can still open up when corrugation exceeds some limiting value. Such
a gap opening does not require any periodic gating correlated with the
corrugation suggested in \cite{low11}. And third, it has been claimed that
one-dimensional corrugations lead to creation of flat bands and a build up
of the density of states (DOS)\ at the Dirac (neutral) point \cite%
{mikmidgap08,guinea}. These partially flat bands are believed to be analogs
of the zero-energy Landau levels because the ripples affect the electrons
like an effective magnetic field. Below, we show that in reality no flat
bands occur since the Dirac points are stable up to the deformation where
the system transforms into a metal.

As in Ref.\onlinecite{mikmidgap08}, we restrict ourselves to the limiting
case of the armchair direction $\boldsymbol{\lambda }=(n,n)$ with the
inversion symmetry being preserved. Such a corrugation has a period of $%
2n\,b_{0}$, where $b_{0}=\sqrt{3}a/2$ (the unit used in Ref.~%
\onlinecite{mikmidgap08}). When $A/\lambda =0$, there are $2n $ conduction $%
\pi^{\ast }$ bands and $2n$ valence $\pi $ bands in the $\Gamma - X$
direction perpendicular to $\boldsymbol{\lambda }$ and containing a Dirac
point (in the supercell BZ) \cite{saito}. Of these $2n$ bands (conduction or
valence), two are nondegenerate and $n-1$ are doubly degenerate \cite{saito}%
. The $\pi ^{\ast } $ and $\pi $ bands that form a Dirac cone are always
nondegenerate (except for the Dirac point itself). When the corrugation is
induced ($A/\lambda \neq 0$), all the doubly degenerate bands split due to
breaking of the mirror symmetry. As $A/\lambda $ is further increased, the
lowest in energy split $\pi^{\ast }$ band and the highest in energy split $%
\pi $ band start moving toward the Fermi level, due to a change in the $\pi
_{z}-\pi _{z}$ hopping matrix elements and increasing $\sigma ^{\ast }-\pi
^{\ast }$ and $\sigma -\pi $ hybridization.

\begin{figure}[h]
\begin{centering}
\includegraphics
[width=8.8cm]{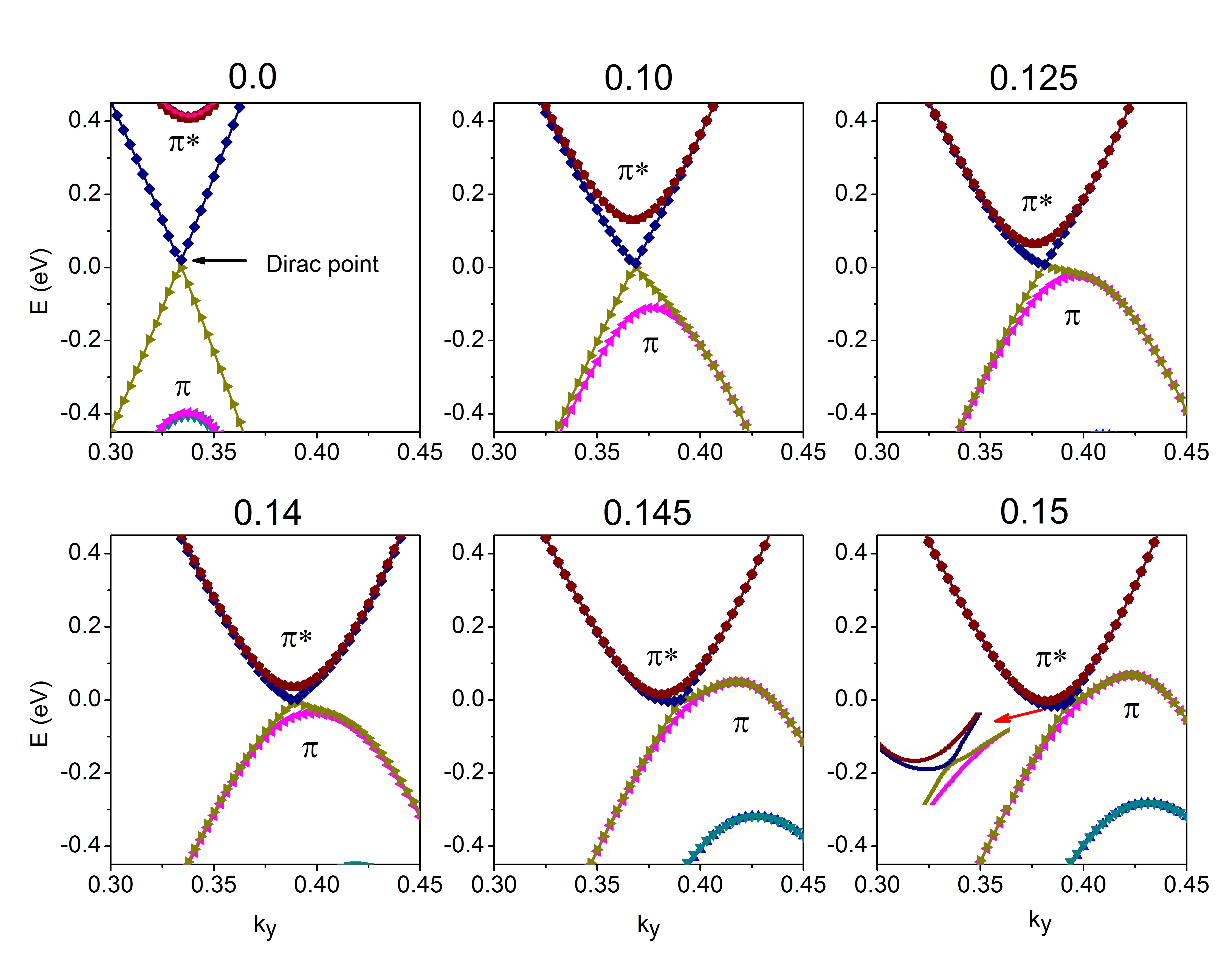}
\caption{(color online).
Bands in graphene corrugated along the
armchair direction with $\boldsymbol{\lambda }=(20,20)$, with amplitude to
period ratio $A/\lambda =0 - 0.15$ and the phase $\varphi =0$ preserving the
inversion symmetry. The $k_{y}$ direction is perpendicular to the rippling
direction. Note that the bands are defined in order of energy.}\label{fig:4S}
\end{centering}
\end{figure}

\begin{figure}[h]
\includegraphics
[width=9.0cm]{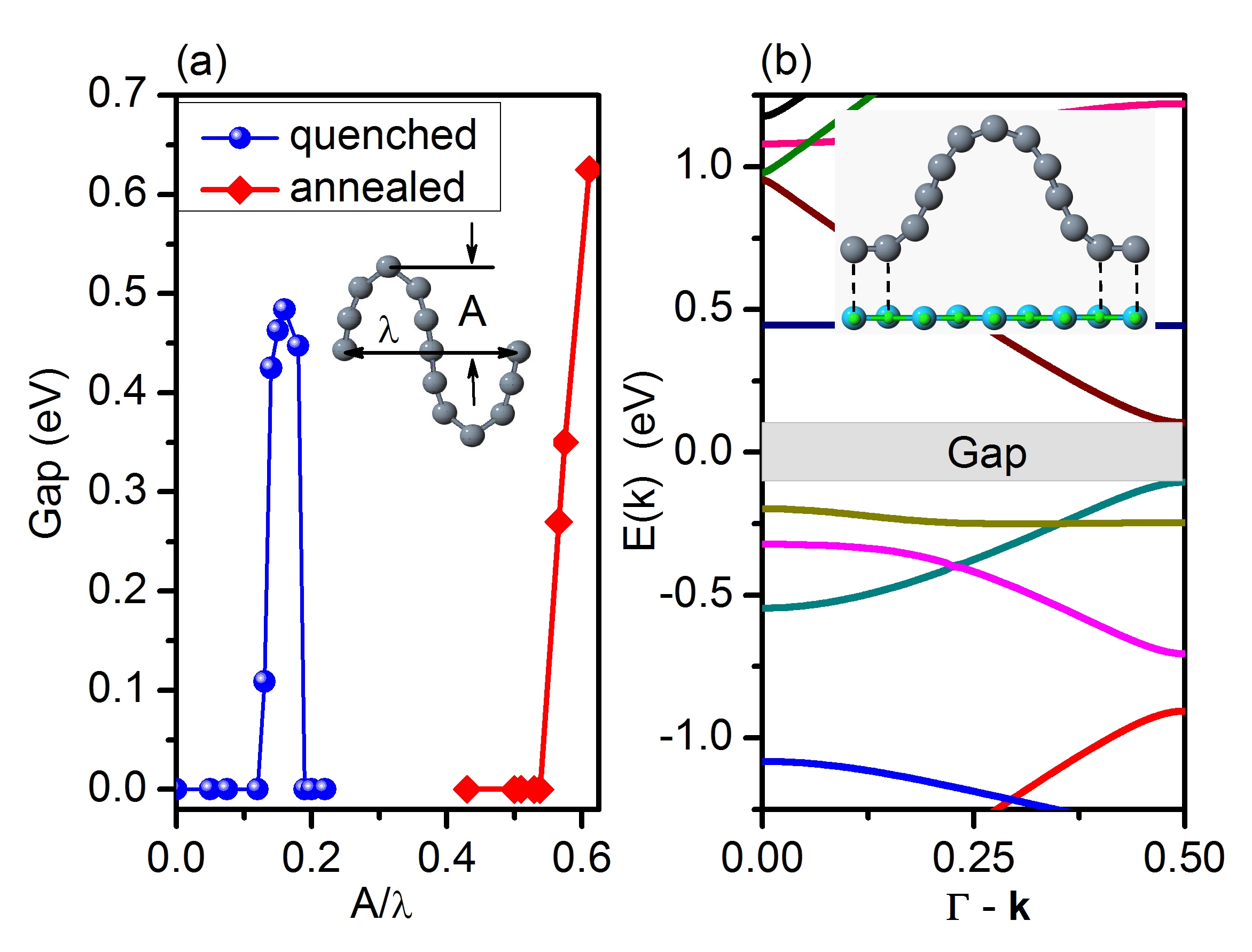} 
\caption{(color online). Band gaps in corrugated periodic graphene sheets.
(a) For quenched and annealed sinusoidal corrugations with $\boldsymbol{%
\protect\lambda}=(6,0)$ as a function of the corrugation parameter $A/%
\protect\lambda$. The critical corrugation ($A/\protect\lambda )_{c}$
significantly increases in passing from the quenched to annealed ripple
structures. The annealed structure at ($A/\protect\lambda )_{c}$ is shown in
the insert. (b) For a corrugated graphene placed on a hexagonal boron
nitride (h-BN) substrate. It is supposed that the connection with the
substrate is only partially coherent: within the period only four graphene
atomic lines are in registry with h-BN lattice (shown by vertical dotted
lines in the insert).}
\label{fig:GonBN}
\end{figure}

To be more specific, consider the ripples with $n=20$ and $\lambda
=40\,b_{0} $, Fig.~\ref{fig:4S}. Here, the closest to the Dirac cone are the
conduction and valence bands residing inside the Dirac cone almost
symmetrically ($A/\lambda =0$, Fig.~\ref{fig:4S}). Such a relative position
of the bands can be easily predicted by using the "band-folding" procedure.
As the $A/\lambda $ increases, the degenerate $\pi ^{\ast }$ and $\pi $
states split and repel each other, so that the lower/upper singly degenerate
$\pi ^{\ast }$/$\pi $ state lowers/rises in energy. By approaching the Dirac
cone, these two states considerably deform the latter. At some critical
moment ($A/\lambda \approx $0.145), the moving $\pi ^{\ast }$ and $\pi $
bands reach the Fermi level leading to a peak in the DOS at the Fermi level
(due to the evident van Hove singularities). In Ref.~\onlinecite{mikmidgap08}%
, this situation was interpreted as a formation of a flat pseudo-Landau
level. In reality, however, we face here a peculiar \emph{electronic
topological transition} of the type semimetal-metal when the bottom of the $%
\pi^\ast$ band and the top of the $\pi $ band pass the Fermi level
simultaneously. It is clear that such a transition is accompanied by
electrons flowing from the $\pi $ to the $\pi ^{\ast }$ band.

So far, we considered sine-wave deformations with only flexural
(out-of-plane) atomic displacements. Such deformations are accompanied by
the stretching $A^{2}k^{2}/4=\left( \pi A/\lambda \right) ^{2}$ which can be
relieved by allowing the atoms to relax at a given amplitude $A$.
`Annealing' increases the initial quenched ratio $A/\lambda $ because $%
\lambda $ shrinks trying to restore the initial nearest-neighbor spacing.
Since, on the other hand, the pace of movement of the Dirac points with $A$
in the relaxed structures is slower, the critical corrugation ($A/\lambda
)_{c}$ for the gap opening becomes considerably higher. For example, the
annealing of the corrugated structure $\boldsymbol{\lambda }=(6,0)$
increases the critical corrugation from 0.12 to $\sim $ 0.5, Fig.~\ref%
{fig:GonBN}a. It is important that in the relaxed structures, the $\sigma
^{\ast }$ band is affected significantly less than in the case of a
stretched lattice and, therefore, one may open up a wider gap ($\sim 1$ eV).

\section{Discussion and conclusions}

It is clear from the above that the periodicity of the corrugations must not
necessarily be commensurate with the undistorted graphene lattice or even
sine-like for the gap opening. The commensurability between the deformation
and the lattice itself would be important if the energy gap were due to
mixing of electronic states belonging to two different valleys $\boldsymbol{K%
}$ and $\boldsymbol{K}^{\prime }$. However, in the case of 1D periodic
corrugations this does not occur even if the corrugations provide a momentum
transfer of $\boldsymbol{k}=\boldsymbol{K-K}^{\prime }$ (as in Fig.1) The
gap opens only due to (i) breaking of the inversion symmetry and/or (ii)
merging of two inequivalent Dirac points, as explained above. We simulated,
for example, the situation when a corrugated graphene is placed on a
hexagonal boron nitride (h-BN) and only a fraction of its atomic lines are
in registry with h-BN. Such a partially coherent connection with the
substrate maintains Gaussian-like periodic corrugations that we may call a
`mesh' (see the insert in Fig.~\ref{fig:GonBN}b). In Fig.~\ref{fig:GonBN}b,
we present the results for a particular case when the graphene is deformed
in such a way that its initial period $\boldsymbol{\lambda }=(6,0)$ becomes
commensurate with the BN period $\boldsymbol{\lambda }=(4,0)$. We see that
the corrugation leads to a gap of 0.20 eV. \ In practice, the periodic
deformations can be achieved in various ways. One can imagine a situation
when the edges are not clamped and a graphene membrane bends in
accordion-like fashion. This situation corresponds to the `annealed'
corrugations where the initial nearest-neighbor lengths are mainly
preserved. The `quenched' corrugations can be induced by depositing graphene
on a substrate like SiO$_{2}$ \cite{ishigami} or the vicinal surfaces with
regularly spaced steps, like Au(788) \cite{weiss}. In the case of SiO$_{2}$,
for example, the interaction energy between the graphene sheet and the
substrate is sufficient to overcome the elastic energy needed for graphene
to conform to the SiO$_{2}$ surface profile \cite{ishigami}.

In conclusion, we have shown that any practical \emph{homogeneous }
deformation cannot open up a gap in graphene sheets. At the same time, the
\emph{inhomogeneous} deformation can open up a significant gap ($\sim $ 1
eV) rather independently of direction and form of a corrugation with an
exception of an armchair direction. The present gap opening does not require
any periodic gating correlated with the corrugation\cite{low11} either. It
is worth noting that in the limit of a clean undoped graphene sheet, the
Coulomb interaction between electron and hole carriers may become
significant if the Coulomb coupling constant exceeds some critical value on
the order of unity and may facilitate an excitonic gap in the spectrum \cite%
{khvesh09}. However, in practical cases the Coulomb interaction may be
screened by (unintentional)\ doping. Since a multilayer graphene
(MLG)/graphite is a stack of weakly coupled graphene sheets \cite{kopel07},
most of above results, therefore, should apply to the inhomogeneously
deformed MLG as well.


\begin{thebibliography}{99}
\bibitem{note1} Results of the gap opening are not affected by this
supercell approximation, as illustrated by gap opening in corrugated
graphene partly attached to the BN substrate, Fig.~\ref{fig:GonBN}b.
\end{thebibliography}
\end{document}